\documentclass{book}    
\usepackage{piers}  
\usepackage{cite}

\begin{document}

\title{Plasmonically Enhanced Spectrally-Sensitive Coatings for Gradient Heat Flux Sensors}
\maketitle

\author      {Kevin M. Conley}
\affiliation {Aalto University}
\address     {}
\city        {Espoo}
\postalcode  {}
\country     {Finland}
\phone       {}    
\fax         {}    
\email       {kevin.conley@aalto.fi}  
\misc        { }  
\nomakeauthor
\author      {Vaibhav Thakore}
\affiliation {University of Western Ontario}
\address     {}
\city        {London}
\postalcode  {}
\country     {Canada}
\phone       {}    
\fax         {}    
\email       {vaibhav.thakore@aalto.fi}  
\misc        { }  
\nomakeauthor
\author      {Tapio Ala-Nissila}
\affiliation {Aalto University}
\address     {}
\city        {Espoo}
\postalcode  {}
\country     {Finland}
\phone       {}    
\fax         {}    
\email       {tapio.ala-nissila@aalto.fi}  
\misc        { }  
\nomakeauthor

\begin{authors}

{\bf Kevin Conley}$^{1}$, {\bf Vaibhav Thakore}$^{2}$, {\bf and Tapio Ala-Nissila}$^{1,3}$\\
\medskip
$^{1}$Department of Applied Physics and QTF Center of Excellence\\
Aalto University, P.O. Box 13500, FI-00076, Aalto, Finland\\
$^{2}$Department of Applied Mathematics\\ 
Western University, London, Ontario N6A 5B7, Canada\\
$^{3}$Interdisciplinary Centre for Mathematical Modelling \\
Departments of Mathematical Sciences and Physics \\
Loughborough University, Loughborough, Leicestershire LE11 3TU, United Kingdom

\end{authors}

\begin{paper}

\begin{piersabstract}
The spectral response and directional scattering of semiconductor-oxide core-shell spherical microparticles embedded in an insulating medium at low volume fraction are computed using Mie Theory and Multiscale Modelling methods. The surface plasmon resonances of low-bandgap semiconductor microinclusions have excellent and tunable scattering properties. By adjusting the size, material, shell thickness, and dielectric environment of the particles, the energies of the localized surface resonances are tuned to match the discrete solar spectrum. Near-IR solar reflectance efficiency factors of up to 78\% are observed. Further the transmittance of broadband or specific wavelengths could be blocked. These spectrally-sensitive coatings have application as a back-reflector for solar devices, high temperature thermal insulator, and optical filters in Gradient Heat Flux Sensors (GHFS) for fire safety applications. 

\end{piersabstract}

\psection{Introduction}
Controlling the propagation of electromagnetic waves using plasmon amplifiers has many applications for metamaterials~\cite{hess2012active}, optical sensors~\cite{kabashin2009plasmonic}, communications~\cite{aydin2011broadband}, and solar cells~\cite{siebentritt2017chalcopyrite}. The drive to improve the efficiency of solar cells has led to the development of ultrathin solar cells. While thin absorber layers reduce the cost of the material, the short circuit current decreases in ultrathin films~\cite{lundberg2003influence}. This loss can be mitigated with a plasmonically-enhanced back-reflector layer which reflects light back into the device~\cite{siebentritt2017chalcopyrite,yin2015integration,van2015light}. The back-reflector layers are embedded with small plasmonic particles with large scattering cross-sections and localized surface plasmon resonances. The plasmon resonance energies can be adjusted by changing the geometry, size, shape, bandgap, and dielectric environment of the nano- and micro-inclusions~\cite{laaksonen2013influence,petryayeva2011localized}. Previously, we have shown low-bandgap semiconductor microinclusions have excellent and customizable near-IR reflectance properties~\cite{tang2017plasmonically}. 

Optically-sensitive coatings and films are also useful for Gradient Heat Flux Sensors (GHFS) which detect near-IR radiation using the transverse Seebeck effect~\cite{mityakov2012gradient}. Coating the sensors with semiconductor-embedded composites selectively reflects unwanted wavelengths and allows the flame radiation to be distinguished from other background sources. Unlike conventional detectors which detect the consequence of fires, such as elevated temperature and smoke, these mm-sized sensors directly measure the heat flux~\cite{mityakov2012gradient}. GHFS have a response time of 10 ns and do not require external power or cooling~\cite{mityakov2012gradient}. This enables the fast detection of fires in critical time-sensitive applications, and provides more information on the ignition stages than current sensors. 

Here we investigate the plasmon resonances of semiconductor microparticles for spectrally-sensitive coatings. We have simulated the spectral response and directional scattering of core-shell microspheres embedded in an insulating medium at low volume fraction (1\%) using Mie theory and Monte Carlo methods. The focus is on the near-IR solar spectra, but the results can be generalized to other sources, including flame spectra for fire safety applications and blackbody radiation for high temperature insulators. 

\psection{Methods}\label{methods}
We consider the dielectric response in an incident electromagnetic field of small semiconductor-oxide core-shell spheres of total radius \textit{R}, core radius \textit{r}, and shell thickness \textit{t} = \textit{R} - \textit{r} as seen in Figure~\ref{fig:model}. The spherical particles are surrounded by a non-absorbing insulating medium with constant refractive index 1.0 or 1.5 and irradiated by near-IR light, $\lambda$ = 1.4 to 4 $\mu$m. The semiconductor (Ge, Si, InP) bandgap ranged from 0.92 to 1.85 $\mu$m, and the oxide (TiO$_2$, SiO$_2$, and ZrO$_2$) relative permittivity spanned from 1.4 to 2.4~\cite{palik1998handbook,wood1982refractive}. 

To consider the effects of the thickness of the oxide coating, the total particle size, \textit{R}, was fixed and the outer semiconductor layer replaced with an oxide with varying thickness, \textit{t}. The filling ratio, $\rho$, is defined as the volume fraction of the oxide within the particle. 

\begin{figure}
\centering
\includegraphics[width=0.3\textwidth]{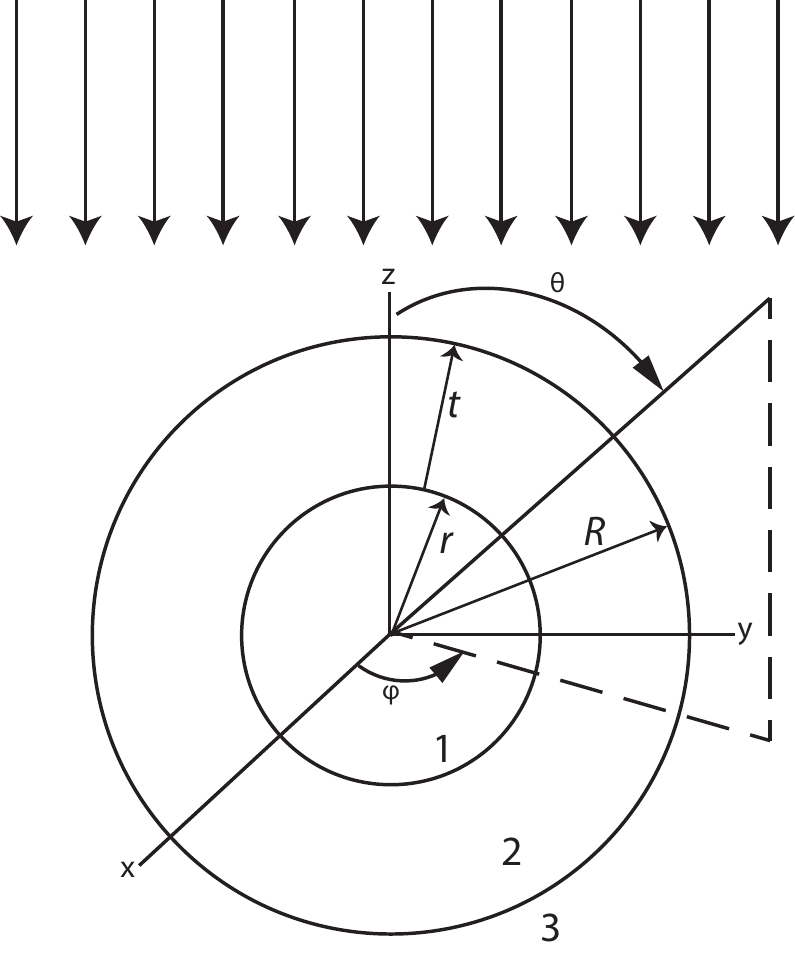}
\caption{\textbf{Model.} Core-shell model adapted from Bohren and Huffman~\cite{bohren2008absorption}. The thickness of the shell, \textit{t}, is the radius of the outer sphere, \textit{R}, minus the radius of the inner sphere, \textit{r}.}
\label{fig:model}
\end{figure}

The Mie coefficients, \textit{a}$_n$ and \textit{b}$_n$, were calculated using a program modified from Bohren and Huffman~\cite{bohren2008absorption}, 

\begin{equation*}
a_n = \frac{\psi_n(y)[\psi'_n(m_2y)-A_n\chi'_n(m_2y)]-m_2\psi'_n(y)[\psi_n(m_2y)-A_n\chi_n(m_2y)]}{\xi_n(y)[\psi'_n(m_2y)-A_n\chi'_n(m_2y)]-m_2\xi'_n(y)[\psi_n(m_2y)-A_n\chi_n(m_2y)]},
\end{equation*}

\noindent and 

\begin{equation*}
A_n = \frac{m_2\psi_n(m_2 x)\psi'_n(m_1 x) - m_1\psi'_n(m_2 x)\psi_n(m_1 x)}{m_2\chi_n(m_2 x)\psi'_n(m_1 x) - m_1 \chi'_n(m_2 x) \psi_n(m_1 x)},
\end{equation*}

\noindent where \textit{m}$_1$ and \textit{m}$_2$ are the refractive indices of the core and shell relative to the surrounding medium and \textit{x} = \textit{kr}, \textit{y} = \textit{kR}, and similarly for \textit{b}$_n$, \textit{B}$_n$. In the limit of zero core radius, \textit{a}$_n$ and \textit{b}$_n$ reduce to those for a homogeneous sphere and lim$_{r\rightarrow 0}$ \textit{A}$_n$ = lim$_{r\rightarrow 0}$ \textit{B}$_n$ = 0. 

The single particle efficiencies of scattering ($Q_{\textrm{sca}}$), absorption ($Q_{\textrm{abs}}$), and extinction ($Q_{\textrm{ext}}$) were calculated as

\begin{equation}
Q_{\textrm{ext}} = Q_{\textrm{sca}} + Q_{\textrm{abs}},
\end{equation}

\begin{equation}
Q_{\textrm{sca}} = \frac{2}{y^2} \sum_{n=1}^N (2n+1)(\lvert a_n \lvert^2 + \lvert b_n \lvert^2),
\end{equation}

\begin{equation}
Q_{\textrm{abs}} = \frac{2}{y^2} \sum_{n=1}^N (2n+1)[\textrm{Re}(a_n + b_n) - (\lvert a_n \lvert^2 + \lvert b_n \lvert^2)],
\end{equation}

\noindent where $y=\frac{2\pi R n_{\textrm{med}}}{\lambda}$ is a size parameter and the summations are truncated at \textit{N} $>$ $y+4y^{1/3}+2$. The particle scattering anisotropy or asymmetry factor, \textit{g}, is given by

\begin{equation}
g = \frac{4}{y^2} \sum_{n=1}^N \big[ \frac{n(n+2)}{n+1} \textrm{Re}(a_n a^*_{n+1} + b_n b^*_{n+1} ) + \frac{2n+1}{n(n+1)} \textrm{Re}(a_n b_n^*) \big].
\end{equation}

The microparticles were well-dispersed in a non-absorbing insulating matrix with refractive index 1.0 or 1.5 with a volume fraction of 0.01. The composite layer thickness, \textit{T}, was 200 $\mu$m, and is surrounded by a non-absorbing ambient medium. The scattering and absorption coefficients, $\mu_{\textrm{sca}}$ and $\mu_{\textrm{abs}}$, of the composite are

\begin{equation}
\mu_{\textrm{sca,abs}} = \frac{3}{2} \frac{fQ_{\textrm{sca,abs}}}{2R},
\end{equation}

\noindent and the effective dielectric permittivity, $\epsilon_{\textrm{eff}}$, of a medium with embedded core-shell spheres from Maxwell Garnett Effective Medium Theory is

\begin{equation*}
\frac{\epsilon_{\textrm{eff}}-\epsilon_\textrm{m}}{\epsilon_{\textrm{eff}}+2\epsilon_\textrm{m}} = f_1\frac{\epsilon_1 - \epsilon_\textrm{m}}{\epsilon_1 + 2\epsilon_\textrm{m}} + f_2\frac{\epsilon_2 - \epsilon_\textrm{m}}{\epsilon_2 + 2\epsilon_\textrm{m}},
\end{equation*}

\noindent and $\epsilon_{\textrm{eff}}$ reduces to uncoated sphere when \textit{f}$_1$ = 0, \textit{f}$_2$ = 0, $\epsilon_1$ = $\epsilon_2$, or $\epsilon_1$ = $\epsilon_\textrm{m}$.

The transmittance, reflectance, and absorbance of a free-standing composite layer were simulated using a Monte Carlo method originally developed by Wang et al.~\cite{wang1995mcml}. The Monte Carlo method records the path and termination result of 10$^7$ photons from an infinitesimally thin beam normal to the composite surface. The grid resolution of \textit{dz} = 0.1 $\mu$m and \textit{dr} = 5 $\mu$m was used for the radial and axial direction, respectively. The total number of grid elements in the axial direction was 100, and angular dependence ignored. The diffuse reflectance and transmittance go to zero as a function of the radius of the layer. The core-shell method was verified against coated nanoparticles~\cite{laaksonen2013influence} and bare microparticles in the small-core limit~\cite{tang2017plasmonically}. 

We define the solar efficiency factor, $\eta$, as

\begin{equation}
\eta = \frac{\int_{\lambda_0}^{\lambda_1} \mathbb{R}(\lambda)I(\lambda) d\lambda}{\int_{\lambda_0}^{\lambda_1} I(\lambda) d\lambda},
\end{equation}

\noindent where $\mathbb{R}$ is the reflectance and \textit{I} is the irradiance corresponding to the spectral density of the electromagnetic radiation emitted by a solar radiation standard~\cite{standard1998g159}. 

\psection{Results and Discussion}\label{Results}

The optical behavior of semiconductor-embedded microcomposite films under near-IR solar radiation was simulated using a Monte Carlo method. The proportion of reflected, absorbed, and transmitted photons by thin films embedded with InP-ZrO$_2$ oxide-coated semiconductors or ZrO$_2$-InP semiconductor shells with \textit{R} = 0.6, \textit{t} = 0.1 $\mu$m are seen in Figure~\ref{fig:eff_cartoon}. The total back-reflectance of the films is high and there is negligible absorbance. Such films are suitable as a filter of broadband or specific wavelengths or to reflect light back into a device. The material parameters, such as the particle size and dielectric environment, adjust the optical selectivity, and will be discussed in detail below. 

\begin{figure}
\centering
\includegraphics[width=\linewidth]{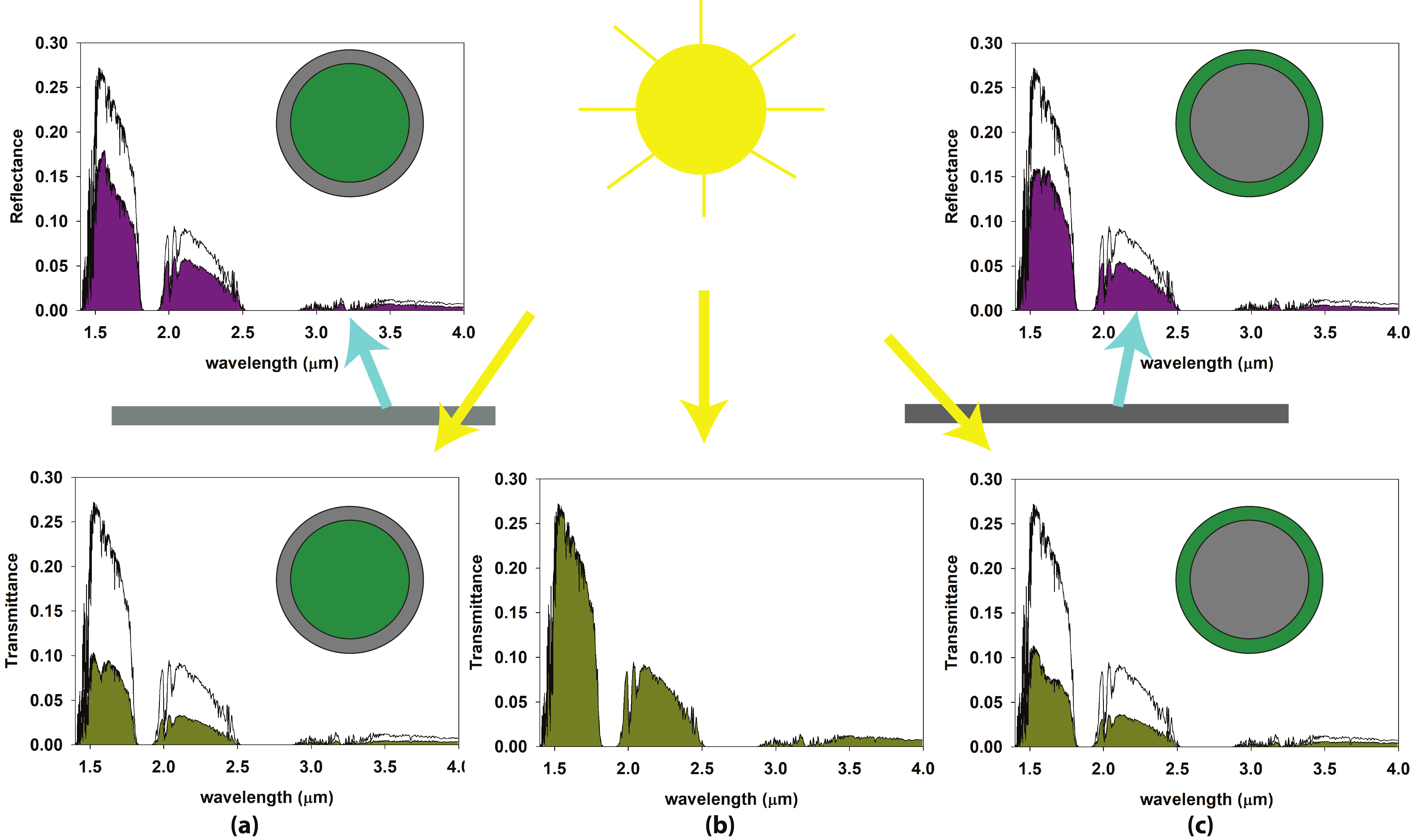}\\
\caption{\textbf{Solar Spectrum Cartoon.} Left - Reflectance and transmittance of a composite of oxide-coated InP-ZrO$_2$ particles, Center - no composite, Right - Composite of ZrO$_2$-InP particles semiconductor shell. The particles (\textit{R} = 0.6 $\mu$m, \textit{t} = 0.1 $\mu$m) are well-dispersed in a non-absorbing, insulating medium with a refractive index of medium is 1.5 and volume fraction of 0.01.}
\label{fig:eff_cartoon}
\end{figure}

To understand how the core and shell sizes affect the film's solar reflectance efficiency factor, $\eta$, the filling ratio, $\rho$, was scanned from 0 to 1. At $\rho$ = 0 or 1, the reflectance efficiency is equivalent to a bare sphere of the shell or core material, respectively. Since the surface resonances of the bare spherical semiconductor also shift with a change in particle size~\cite{tang2017plasmonically}, the efficiency at $\rho$ = 0 changes for different particle sizes as seen in Figure~\ref{fig:eff_scan}(c). For example, the solar reflectance efficiency of InP with \textit{R} = 0.6 $\mu$m is 62\%, but at \textit{R} = 1.0 $\mu$m, the efficiency is only 49\%. The scattering from oxides at $\rho$ = 1, is not a plasmonic effect, and there is no particle size effect.

\begin{figure}
\centering
\includegraphics[width=0.45\linewidth]{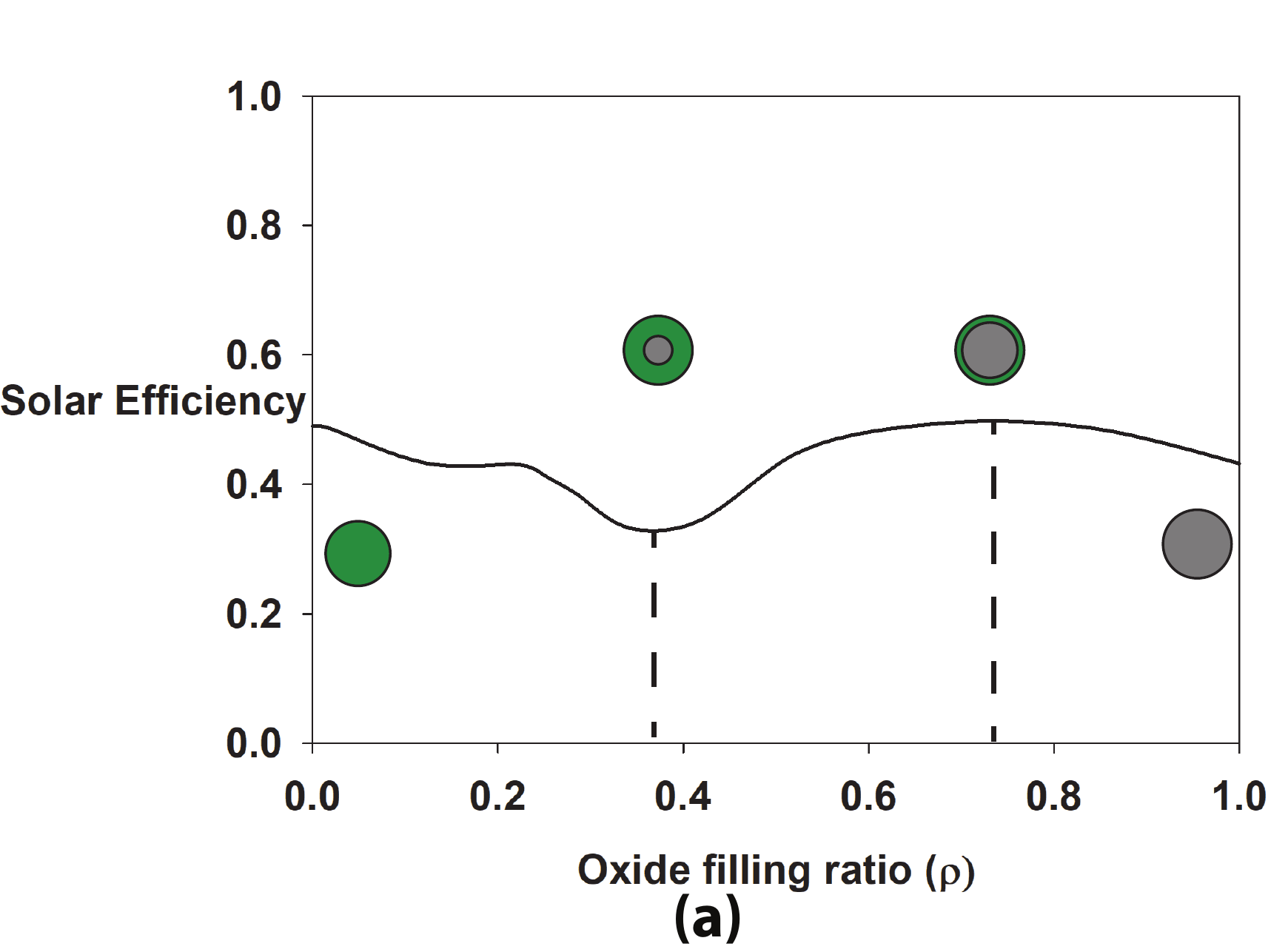}
\includegraphics[width=0.45\linewidth]{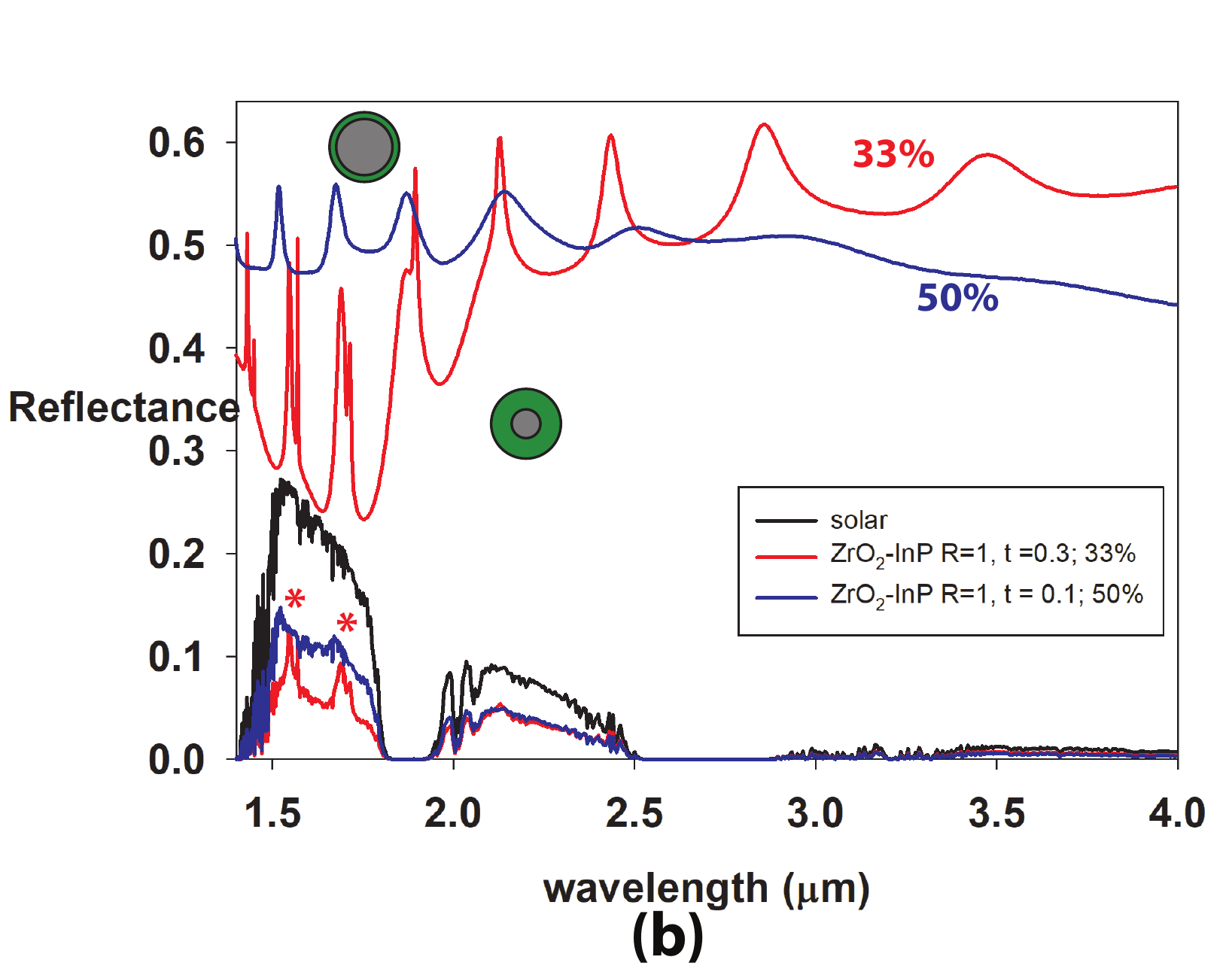}
\\
\includegraphics[width=.45\linewidth]{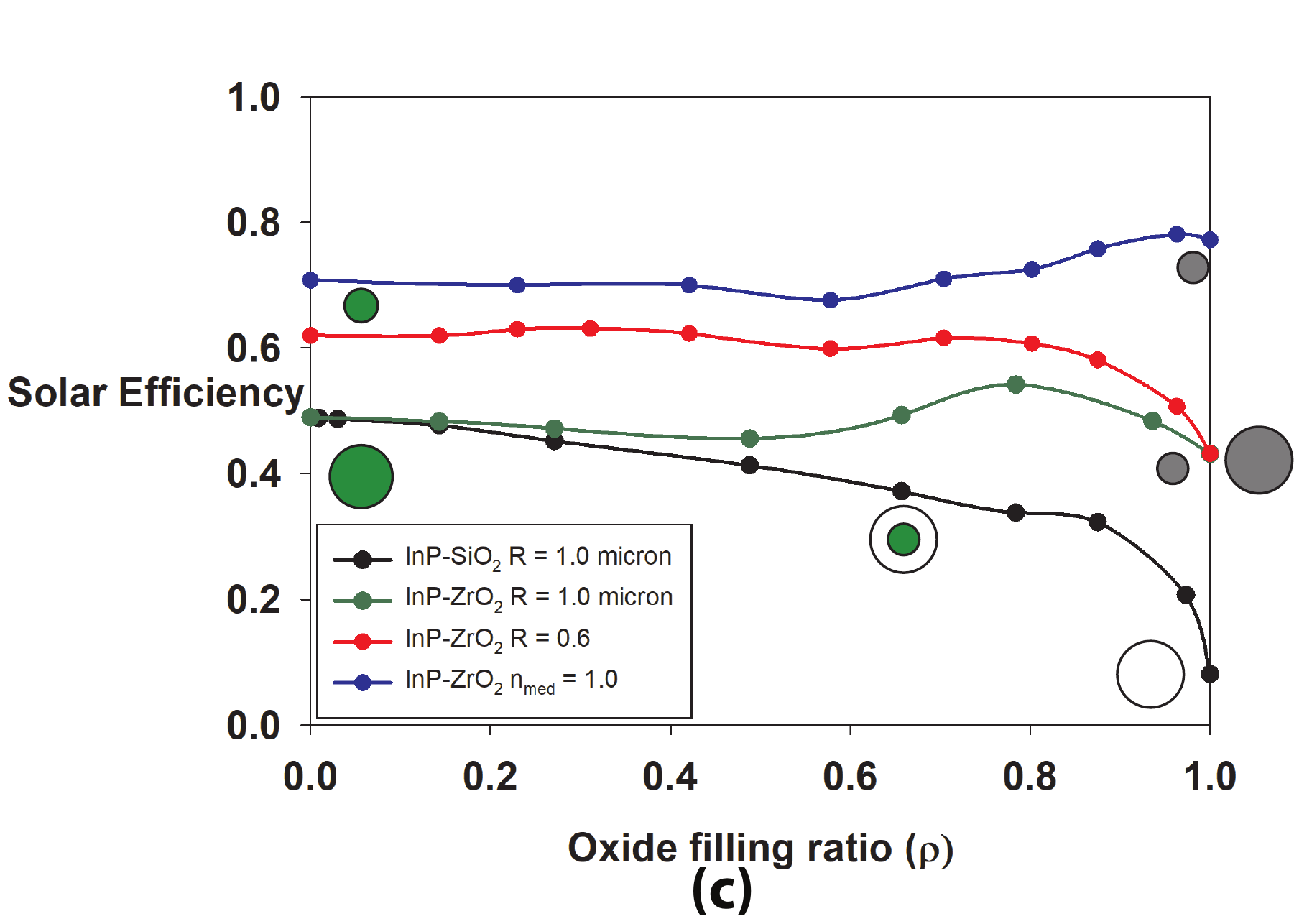}
\caption{\textbf{Solar Efficiency Scan.} Top left - Solar efficiency of ZrO$_2$-InP core-shell particles (\textit{R} = 1.0 $\mu$m) plotted against the volume fraction of the oxide. Top right - Reflectance spectra and simulated solar reflectance spectra of ZrO$_2$-InP at $\rho$ with maximum and minumum solar efficiancy. These correspond to shell thicknesses of 0.1 and 0.3 $\mu$m, respectively. Bottom - InP semiconductors coated with ZrO$_2$ or SiO$_2$ (\textit{R} = 1.0~or 0.6 $\mu$m). The refractive index of medium is 1.5 unless otherwise indicated.}
\label{fig:eff_scan}
\end{figure}

Sweeping $\rho$ leads to local maxima and minima in the reflectance efficiency as shown in Figure~\ref{fig:eff_scan}. When the strongly scattering surface plasmon resonances align with the discrete peaks in the near-IR solar spectra, the total solar reflectance efficiency factor increases. For example composite films embedded with ZrO$_2$-InP particles with \textit{R} = 1.0 $\mu$m have a local maximum at $\rho$ = 0.75 and a local minimum at $\rho$ = 0.35 as seen in Figure~\ref{fig:eff_scan}(a). The maximum, corresponding to a shell thickness of 0.1 $\mu$m, has high, broad reflectance across the near-IR and a solar reflectance efficiency of 50\% as seen in Figure~\ref{fig:eff_scan}(b). At the minimum, $\rho$ = 0.45 and the shell thickness is 0.3 $\mu$m, and the surface plasmon resonance is no longer as broad and the reflectance efficiency factor decreases to 33\%. Although these resonances do not reflect broadly in the solar spectrum, there are, however, sharp peaks with locally high reflectance. High sharp reflectance peaks are observed at 1.55 and 1.69 $\mu$m in the corresponding solar reflectance spectrum of composite films with ZrO$_2$-InP and a shell thickness of 0.3 $\mu$m as seen in Figure~\ref{fig:eff_scan}(b). 

\begin{figure}
\centering
\includegraphics[width=0.49\textwidth]{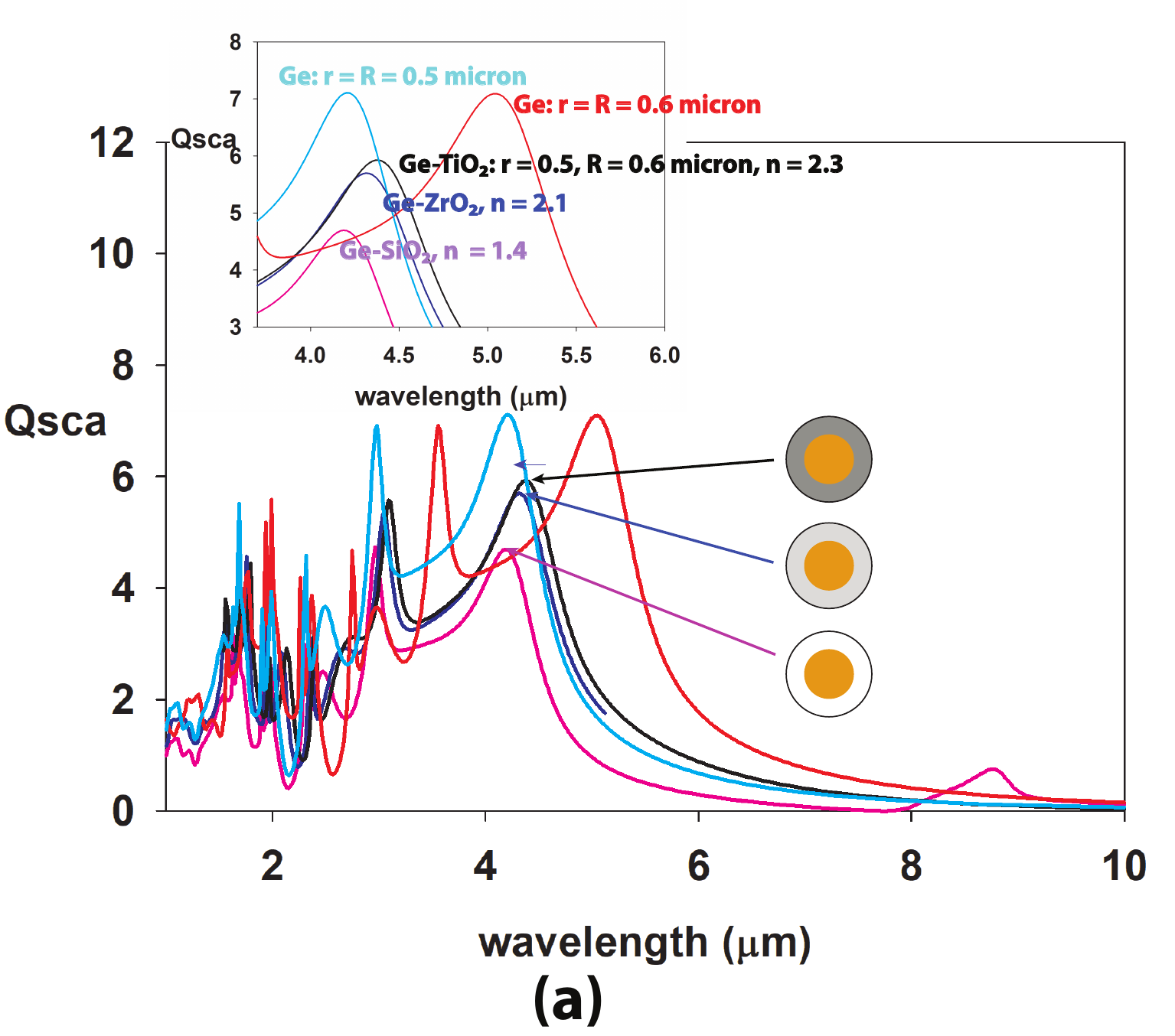}
\includegraphics[width=0.49\textwidth]{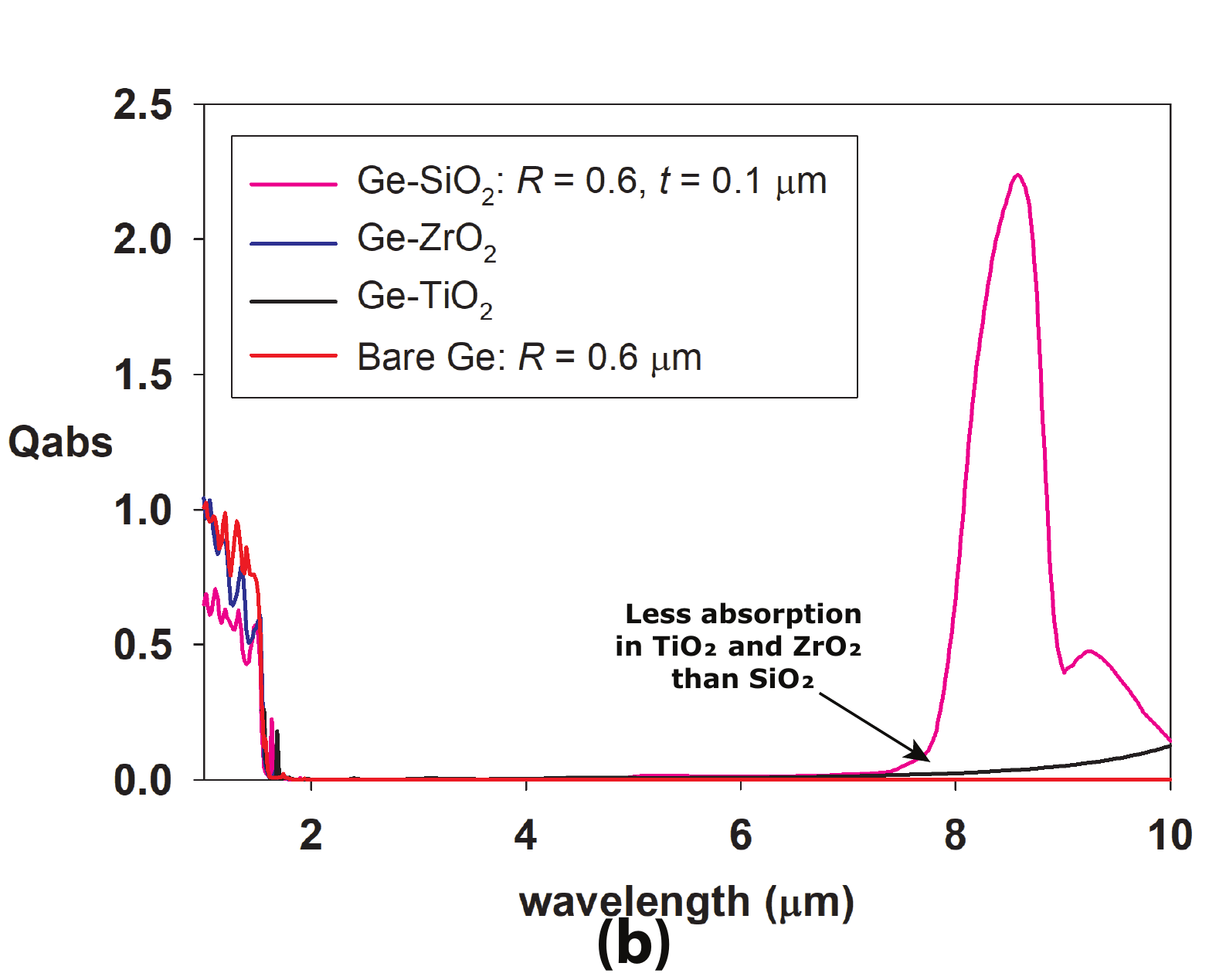}
\caption{\textbf{Oxide effect.} Single particle efficiencies of Ge microparticles, \textit{R} = 0.6 $\mu$m, coated with TiO$_2$, SiO$_2$, and ZrO$_2$.}
\label{fig:Qoxide}
\end{figure}

The presence of oxide in the core-shell particles also affects the absorption and plasmon energy. Core-shell particles with oxide component often absorb light at longer wavelengths due to the large extinction coefficient of the oxide. SiO$_2$, for example, has Si-O stretching modes between 8 - 10 $\mu$m. Thus, microparticles coated with SiO$_2$ have high absorption at wavelengths greater than 8 $\mu$m as seen in Figure~\ref{fig:Qoxide}. As the volume fraction of the oxide increases the absorption efficiency becomes stronger, and can influence the broadband reflectance at longer wavelengths. Other oxides absorb less strongly. Oxide absorption is not relevant to this paper. 

When the refractive index of the shell is close to that of the medium, the core-shell Mie coefficients collapse to that of a bare particle and the shell effectively disappears. This effect is apparent in the single particle scattering efficiencies of oxide-coated Ge (\textit{R} = 0.6 $\mu$m) in Figure~\ref{fig:Qoxide}(a). Of the oxides considered here, SiO$_2$ (\textit{n} = 1.45) is the closest to the refractive index of the medium of 1.5. Thus the energy of the resonances are equivalent to a bare sphere with the same core radius (\textit{r} = 0.5 $\mu$m) as seen in Figure~\ref{fig:Qoxide}(a). In the coated case, the total particle size is larger and the magnitude of the scattering efficiency, $Q_{\textrm{sca}}$, is dampened compared to the bare particle. ZrO$_2$ (\textit{n} = 2.1) and TiO$_2$ (\textit{n} = 2.3) are closer to the refractive index of the semiconductor (\textit{n} = 4). Thus the scattering efficiencies are larger and shifted closer to the bare Ge particle without any peeling (\textit{R}= 0.6 $\mu$m). 

The different oxide shells shift the plasmon resonance and the scattering efficiency to various extents. These shifts change the alignment with the discrete solar spectrum. The efficiency factor of particles coated with SiO$_2$ decreases with increased oxide content. For example the solar efficiency factor at high $\rho$ of InP-SiO$_2$ trails off to less than 10\% while InP-ZrO$_2$ remains high as seen in Figure~\ref{fig:eff_scan}(c). The refractive index of SiO$_2$ matches the medium and it does not scatter well. On the other hand, ZrO$_2$ scatters and its efficiency factor is much higher. It is possible to increase the scattering efficiency of the resonances by choosing a shell with a higher refractive index than the core. 

Altering the dielectric environment of the particles is another mechanism of tuning the plasmon resonances~\cite{tang2017plasmonically}. The scattering resonance efficiencies greatly increase when the refractive index of the medium is lowered from 1.5 to 1.0. The solar efficiency of composite of InP-ZrO$_2$ (\textit{R} = 0.6 $\mu$m, \textit{t} = 0.4 $\mu$m) increased from 51\% to 78\% when the medium refractive index is decreased to 1.0.

In many applications, the transmittance of the films is the crucial criteria. For these coatings, competing absorption mechanisms need to be considered. As the number of charge carriers increases at energies above the semiconductor bandgap, absorption becomes efficient. The bandgap of InP is 0.92 $\mu$m and the absorption has been neglible in the cases considered above, but changing the semiconductor material affects the bandgap onset. In Ge particles, the absorption efficiency becomes stronger at 1.85 $\mu$m (0.67 eV) and can be useful for devices. For example, InP spheres (\textit{r} = 0.5 $\mu$m) coated with Ge (\textit{t} = 0.1 $\mu$m) produce a sharp absorbance resonance at $\lambda$ = 1.59 $\mu$m as seen in Figure~\ref{fig:blocked_lambda}(a). Increasing the shell thickness to 0.2 $\mu$m shifts the resonance to $\lambda$ = 1.69 $\mu$m and a second broader absorption state forms at $\lambda$ = 1.63 $\mu$m. Increasing the shell thickness decreases the confinement of the surface states and lowers the energy of the defect state. Characteristic of plasmonic systems, in larger shells there is a greater separation of the oscillating charges and the resonance energy is redshifted. 

\begin{figure}
\centering
\includegraphics[width=.45\linewidth]{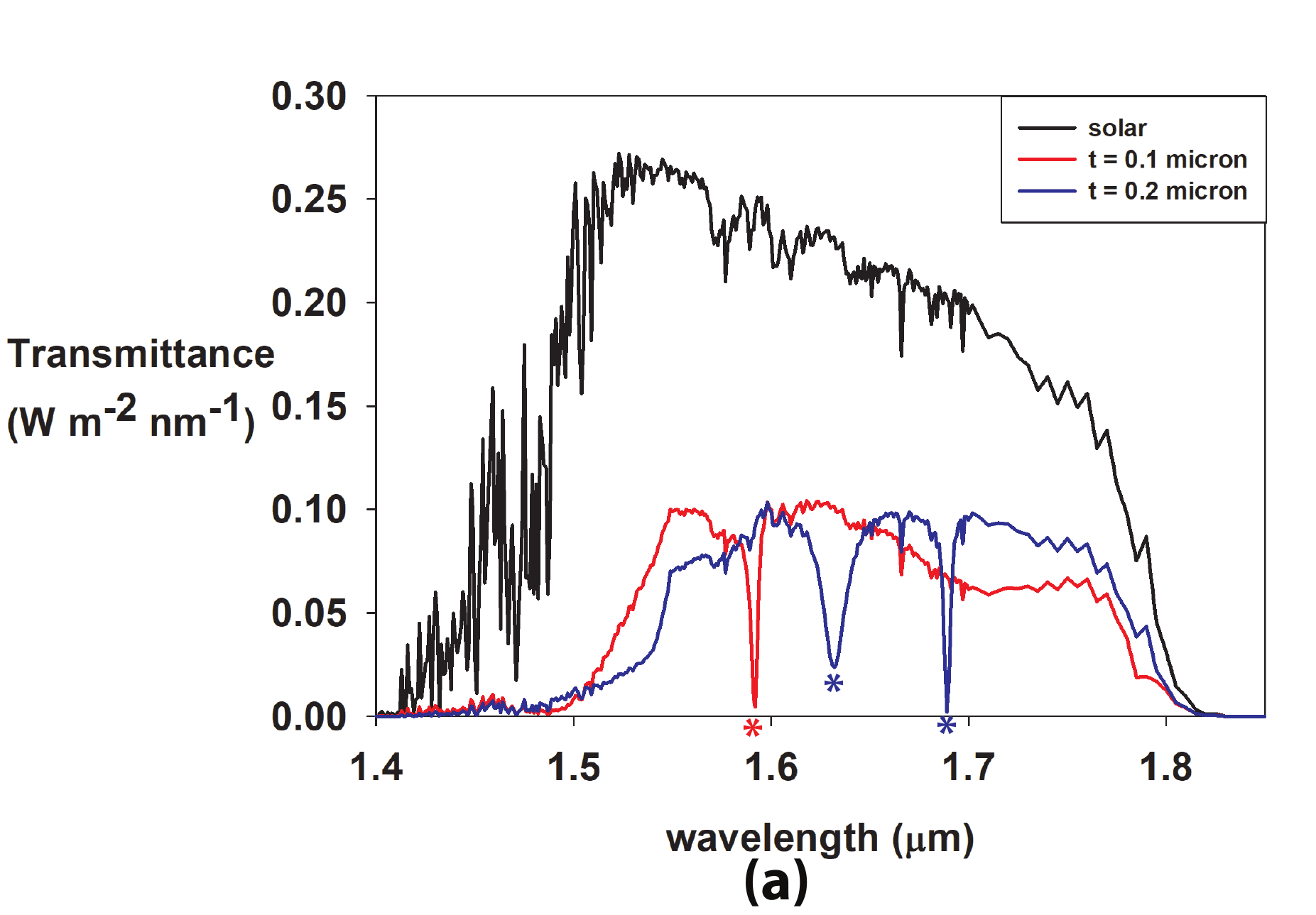}
\includegraphics[width=.45\linewidth]{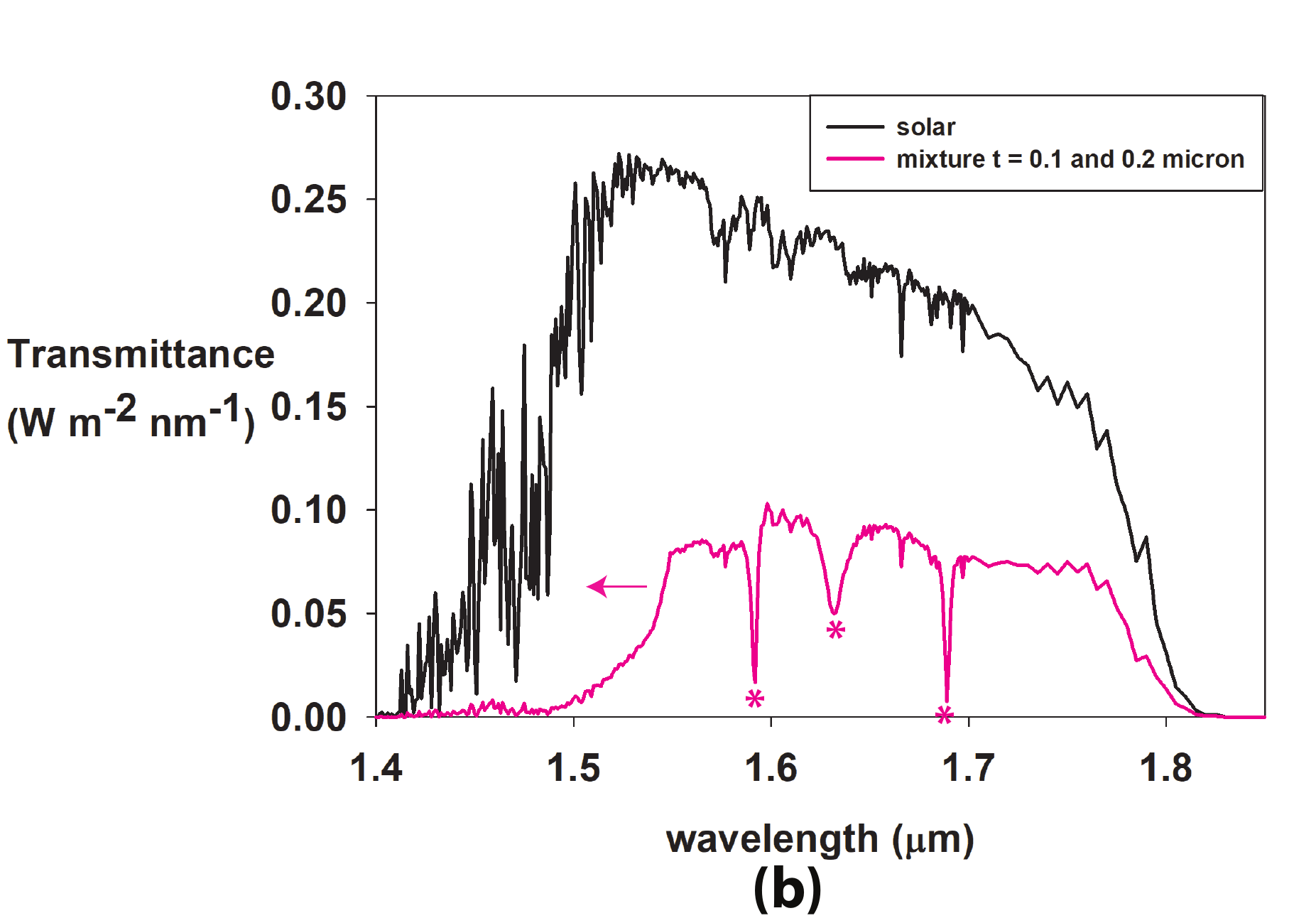}
\caption{\textbf{Plasmon-enhanced blocking.} Left - Transmittance of the primary near-IR solar band through composites of InP-Ge particles (\textit{R} = 0.6, \textit{t} = 0.1 and 0.2 $\mu$m). Right - mixture of the particles with (0.5\% each of \textit{t} = 0.1 and 0.2 $\mu$m) showing blocking of transmittance at $\lambda$ = 1.59, 1.63, and 1.69 $\mu$m (indicated by stars) and broadband blocking of $\lambda$ less than 1.55 $\mu$m (red arrow).}
\label{fig:blocked_lambda}
\end{figure}

Composites with a mixture or distribution of sizes will yield many sharp resonances and further tunability. This enables several wavelengths to be targeted as seen in Figure~\ref{fig:blocked_lambda}(b). Additionally, the small coating of Ge limits the transmittance of light less than 1.55 $\mu$m. The thin Ge layer is enough to generate charge carriers into the Ge conduction band and initiate absorbance. Thicker Ge layers block progressively higher wavelengths up to about 1.6 $\mu$m. 

Despite being moderately forward scattering, composite devices of semiconductor spheres are efficient back reflectors~\cite{tang2017plasmonically}. The directional scattering will not matter in a low density embedded composite layer if the photons can be quickly scattered along random directions. 

\psection{Conclusions}
The highly efficient and tunable scattering cross-section of semiconductor microinclusions allows for the design of plasmonically-enhanced optically-sensitive coatings. The localized surface plasmon resonances of the spherical semiconductors, reflect up to 90\% of the incident light at specific wavelengths. We have simulated the spectral response and directional scattering of coatings embedded with spherical microparticles at low volume fraction using Mie Theory and Monte Carlo methods. By adjusting the particle dimensions, material, and dielectric environment, the energy of the surface plasmon resonances can be tuned to match the incident spectrum. Core-shell spheres provide customization and enhance the total solar reflectance efficiency factor to up to 78\% of the near-IR solar spectrum. Further, the transmittance of specific wavelengths can be blocked by sharp resonances. These coatings are useful as optically-sensitive back-reflectors or transmittance filters for GHFS devices. 

\ack
This work was performed as part of the Academy of Finland Centre of Excellence program (project 312298).

\bibliographystyle{IEEEtran}

\bibliography{PIERS-arxiv}

\end{paper}

\end{document}